\documentclass{achemso}

\usepackage[version=3]{mhchem} 
\usepackage[T1]{fontenc} 
\usepackage{textgreek} 
\usepackage{siunitx} 
\DeclareSIUnit\photon{photon}
\DeclareSIUnit\molar{M}
\DeclareSIUnit\Molar{M}
\DeclareSIUnit\rev{rev} 
\DeclareSIUnit\OD{OD}
\DeclareSIUnit\angstrom{\text{Å}}


\usepackage{soul}

\usepackage[utf8]{inputenc}
\DeclareUnicodeCharacter{2212}{$-$} 

\author{George A. Sutherland}
\affiliation{Plants, Photosynthesis and Soil, School of Biosciences, University of Sheffield, Sheffield S10 2TN, UK}
\altaffiliation{These authors contributed equally.}
\author{James P. Pidgeon} 
\affiliation{Department of Physics and Astronomy, University of Sheffield, Sheffield S3 7RH, UK}
\altaffiliation{These authors contributed equally.}
\author{Harrison Ka Hin Lee}
\affiliation{SPECIFIC, Faculty of Science and Engineering, Swansea University, Swansea SA1 8EN, UK}
\author{Matthew S. Proctor}
\affiliation{Plants, Photosynthesis and Soil, School of Biosciences, University of Sheffield, Sheffield S10 2TN, UK}
\author{Andrew Hitchcock}
\affiliation{Plants, Photosynthesis and Soil, School of Biosciences, University of Sheffield, Sheffield S10 2TN, UK}
\author{Shuangqing Wang} 
\affiliation{Department of Physics and Astronomy, University of Sheffield, Sheffield S3 7RH, UK}
\author{Dimitri Chekulaev} 
\affiliation{Department of Chemistry, University of Sheffield, Sheffield S3 7HF, UK}
\author{Wing Chung Tsoi}
\affiliation{SPECIFIC, Faculty of Science and Engineering, Swansea University, Swansea SA1 8EN, UK}
\author{Matthew P. Johnson}
\affiliation{Plants, Photosynthesis and Soil, School of Biosciences, University of Sheffield, Sheffield S10 2TN, UK}
\author{C. Neil Hunter}
\affiliation{Plants, Photosynthesis and Soil, School of Biosciences, University of Sheffield, Sheffield S10 2TN, UK}
\author{Jenny Clark}
\email{jenny.clark@sheffield.ac.uk}
\phone{+44 (0)114 222 3526}
\affiliation{Department of Physics and Astronomy, University of Sheffield, Sheffield S3 7RH, UK}

\title[V1 - November 2021]
  {Twisted carotenoids do not support efficient intramolecular singlet fission in the orange carotenoid protein}
  
\abbreviations{IR,NMR,UV}
\keywords{singlet exciton fission, singlet fission, orange carotenoid protein, carotenoid, canthaxanthin}

\begin{document}







\pagebreak
\begin{abstract}
Singlet exciton fission is the spin-allowed generation of two triplet electronic excited states from a singlet state. Intramolecular singlet fission has been suggested to occur on individual carotenoid molecules within protein complexes, provided the conjugated backbone is twisted out-of-plane (giving a \textnu\textsubscript{4} $\sim\SI{980}{\per\centi\m}$ resonance Raman peak). However, this hypothesis has only been forwarded in protein complexes containing multiple carotenoids and bacteriochlorophylls in close contact. To test the hypothesis on twisted carotenoids in a `minimal' one-carotenoid system, we study the orange carotenoid protein (OCP). OCP exists in two forms: in its orange form (OCPo), the single bound carotenoid is twisted, whereas in its red form (OCPr), the carotenoid is planar. To enable room-temperature spectroscopy on canthaxanthin-binding OCPo and OCPr without laser-induced photoconversion, we trap them in trehalose glass. Using transient absorption spectroscopy, we show that there is no evidence of long-lived triplet generation through intramolecular singlet fission, despite the canthaxanthin twist in OCPo. 
\end{abstract}

\section{Introduction}

Singlet exciton fission (SF) is the conversion of a spin-0 singlet exciton \cite{Bardeen2014} (or excited singlet state) into a pair of spin-1 triplet excitons \cite{Smith2013, Musser2019a, Ullrich2021, Kim2018}. This multiexciton generation process has been studied over the past decade primarily because of its promise to improve solar cell efficiency \cite{Hanna2006, Rao2017, Ehrler2021, Daiber2021, Ehrler2022}; one-high energy photon creates two low-energy excited states, which could be harvested by conventional photovoltaic devices in a process  minimizing energetic losses due to thermalization. SF has other potential applications for non-linear optics \cite{Liu2015, Zhao2015, Tonami2019}, OLEDs \cite{Nagata2018} or even quantum technologies \cite{Teichen2015,Bardeen2019, Marcus2020a,Smyser2020} by taking advantage of the virtue that a single photon creates a pair of spin-entangled quantum states. However, despite promising results \cite{Einzinger2019a,Daiber2020}, practical applications have yet to be realized, in part due to the limited library of materials that undergo SF, none of which is yet ideal \cite{Rao2017, Ullrich2021}. 

In the search for other SF materials, the polyenes, `class III' SF materials according to Smith and Michl's categorization \cite{Smith2013}, form an intriguing materials class. In these materials, the lowest-lying singlet excited state (S\textsubscript{1}) has dominant triplet-pair character, denoted \textsuperscript{1}(TT) (see Refs.~\cite{Tavan1987, Schmidt2012, Musser2019a}) and thus demonstrates negligible one-photon absorption from the ground-state. S\textsubscript{1} is instead accessed by internal conversion following excitation to the strongly absorbing S\textsubscript{2} state.  

\begin{figure}[ht]
	\centering
	\includegraphics[scale=1.25]{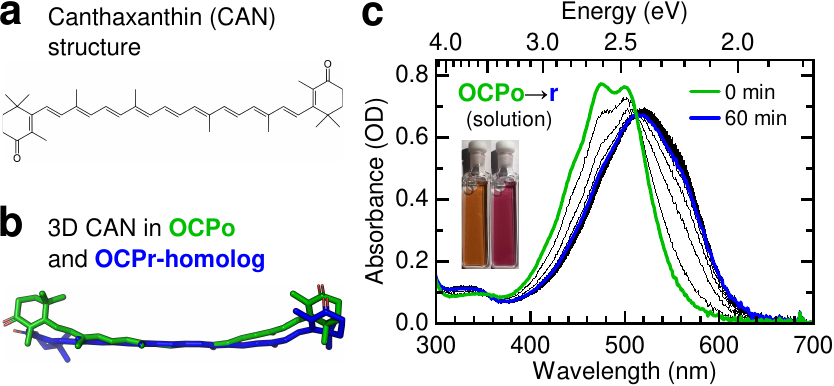}
	\caption{\textbf{The OCP photoswitches from orange (OCPo) to red (OCPr) forms with different carotenoid conformations.} The OCP studied here binds a single CAN carotenoid whose skeletal structure is shown in panel (a). The bound CAN conformation depends on the OCP form, as shown in panel (b): when CAN is bound in OCPo (green, data from X-ray diffraction structure PDB 4XB5 \cite{Leverenz2015, 4XB5}) it has a twisted conformation; when bound in an OCPr N-terminal homolog (blue, data from X-ray diffraction structure of red carotenoid protein, RCP: PDB 4XB4 \cite{Leverenz2015, 4XB4}) it is planar. (c) In solution, OCPo converts to OCPr under white-light illumination (\SI{1600}{\micro\mol\photon\per\square\m\per\s}) resulting in a change in color (see inset) and absorbance spectrum (main panel). The spectra were taken in \SI{1}{\minute} intervals under constant white-light illumination. In the dark, OCPr converts back to OCPo (see SI Fig.~\ref{fig:OCPr_o_solution}). The optical path length for solution measurements was \SI{1}{\milli\m}.} 
	\label{fig:OCP_structures_crts}
\end{figure}

This SF class includes conjugated polymers such as polydiacetylene\cite{Kraabel1998, Pandya2020, Lanzani2001}, poly(alkyl-thienylenevinylene)\cite{Musser2013,Lafalce2011,Musser2019}, a new generation of donor-acceptor singlet fission polymers\cite{Busby2015,Kasai2015,Fallon2019,Huynh2017}, quinoidal thiophenes \cite{Casado2012, Varnavski2015, Chien2015, Kim2018b, Kawata2016}, carbene-based diradicaloids \cite{Ullrich2020} and antiaromatic core-structured molecules \cite{Wu2017, Liu2019c}. The polyene family also includes the carotenoids, a large class of over 1000 naturally occurring molecules,\cite{Canniffe2021,Yabuzaki2017} represented here by canthaxanthin (CAN) which forms the subject of this work (see structure in Fig.~\ref{fig:OCP_structures_crts}a).
\\

In comparison with better-studied `class I' SF materials \cite{Smith2013, Yong2017, Pun2019, Korovina2020, Wang2021}, mostly based on molecules such as pentacene\cite{Wilson2011,Bossanyi2021} or tetracene \cite{Burdett2013, Piland2015a, Tayebjee2013, Wilson2013b}, SF in polyenes is less well understood. This is attributable in part to their complex manifold of low-lying triplet-pair states \cite{Valentine2020, Manawadu2022, Barford2022} and strong vibronic coupling\cite{Balevicius2016, Balevicius2019}, and also partly due to the sensitivity of the photophysics to conjugation length and molecular geometry. In polyenes, the lowest-lying \textsuperscript{1}(TT) state that makes up the dominant contribution of S\textsubscript{1} \cite{Tavan1987, Schmidt2012, Taffet2020, Barford2001} contains tightly-bound triplets that are unlikely to easily separate into free triplets \cite{Polak2019} without additional energy \cite{Valentine2020}. 

Indeed, while intramolecular singlet fission (iSF) has been observed in a variety of long-chain polyenes in solution \cite{Kraabel1998, Pandya2020, Lanzani2001,Musser2013,Lafalce2011,Musser2019,Busby2015,Kasai2015,Fallon2019,Huynh2017}, unlike the recently designed `class I' iSF systems \cite{Wang2021, Korovina2020, Pun2019, Yablon2022}, the triplet-pairs in polyenes decay rapidly (\SI{}{\pico\s}-\SI{}{\nano\s}) to the S\textsubscript{0} ground-state \cite{Musser2019a}. Even in carotenoid aggregates, where intermolecular SF occurs between neighboring chromophores \cite{Musser2015, Zhang2018c,Chang2017, Wang2010,Wang2011, Sutherland2020}, the majority of triplet excited states decay to S\textsubscript{0} surprisingly quickly (within a nanosecond) \cite{Musser2015, Musser2019a, Sutherland2020}. In isolated carotenoids in solution, the dominant deactivation channel from the photoexcited S\textsubscript{2} state is internal conversion to S\textsubscript{1}. To our knowledge, there is no evidence that isolated carotenoids in solution demonstrate intramolecular SF (iSF).

Nevertheless, similarly to recent reports that torsion or twisting along a molecular backbone can allow both rapid iSF and formation of long-lived triplets in `class I' SF materials \cite{Korovina2020,Yablon2022}, iSF along a single \emph{twisted} carotenoid chain to produce long-lived (\SI{}{\micro\s}) triplets has been suggested to occur in some photosynthetic light-harvesting complexes \cite{Gradinaru2001,Papagiannakis2002, Yu2017, Niedzwiedzki2017, Zhang2022}. In these systems, the protein binds the carotenoid so that it is constrained in a twisted geometry. This twist reportedly stabilizes a triplet at either end of the molecule \cite{Gradinaru2001,Yu2017}. 

This hypothesis was initially proposed to explain the presence of SF in the light-harvesting antenna (LH1) from \textit{Rhodospirillum rubrum}, because of the large intermolecular distances between neighboring carotenoids ($>$\SI{10}{\angstrom}) \cite{Gradinaru2001}. More recently, Yu \textit{et al.}\cite{Yu2017}~observed a correlation between the presence of SF and the so-called \textnu\textsubscript{4} resonance Raman peak ($\sim\SI{980}{\per\centi\m}$) in light-harvesting complexes (LH1-RC and LH2) from \textit{Thermo\-chromatium tepi\-dum} and \textit{Rhodobacter sphaeroides} 2.4.1. The intensity of \textnu\textsubscript{4} is related to carotenoid backbone twisting,\cite{Yu2017,Kish2015} so this finding led to the conclusion that backbone twisting of the carotenoid directly enables iSF.

To test the hypothesis that singlet fission (SF) can occur along a single twisted carotenoid chain, we examine a protein that binds a single carotenoid: the orange carotenoid protein (OCP). In OCP the protein exists in two forms, orange and red (OCPo and OCPr) with the carotenoid in either a twisted or planar conformation, respectively (see Fig.~\ref{fig:OCP_structures_crts}). By studying both forms with the protein fixed in a trehalose-sucrose glass, we demonstrate that a twisted backbone is not sufficient to enable iSF in a protein-bound carotenoid. In light of recent work understanding magnetic field effects (MFEs) in SF systems \cite{Bayliss2016, Bossanyi2021b}, we also discuss published reports of MFE in light-harvesting complexes from purple bacteria \cite{Kingma1985a, Kingma1985, Rademaker1980, Klenina2013, Klenina2014, Gryaznov2019} and find that the reported MFEs are also inconsistent with iSF. Overall, we conclude that iSF is not supported on carotenoids bound to the OCP and unlikely to occur in light-harvesting complexes. 

\section{Results}

In this study, the OCP was produced in \textit{Escherichia} (\textit{E.}) \textit{coli} by virtue of a dual plasmid system comprised of pET28a with the \textit{Synechocystis} sp. PCC 6803 OCP gene (slr1963) and pAC-CANTHipi, which provides near-100\% accumulation of CAN.\cite{Cunningham2007} Carotenoid-containing protein was isolated according to the method described in the methods Section.

\begin{figure}[h!]
	\centering
	\includegraphics[scale=1.25]{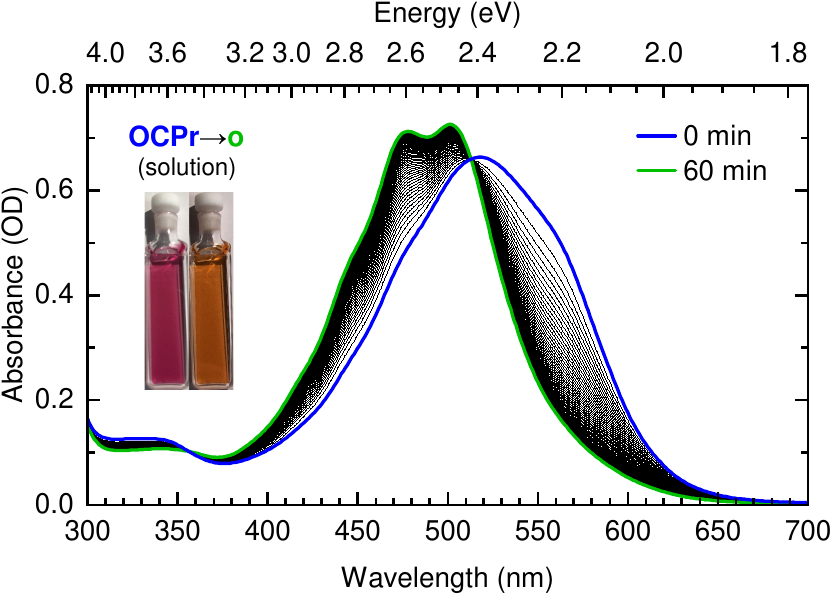}
	\caption{\textbf{Backconversion of OCPr to OCPo in solution.} Steady-state absorbance spectra of OCPr in solution taken in \SI{1}{\minute} intervals at \SI{22}{\degreeCelsius} in darkness. The optical path length was \SI{1}{\milli\m}.}
	\label{fig:OCPr_o_solution}
\end{figure}

In solution, upon illumination with white light, the dark-adapted OCPo form undergoes a conformational switch to the OCPr form, with a concomitant red-shift of its steady-state absorbance spectrum due to the effective conjugation length extension of the bound carotenoid \cite{Wilson2008, Niedzwiedzki2014, Kish2015, Bondanza2020}, see Fig.~\ref{fig:OCP_structures_crts}c. The change is reversible, with back-conversion from OCPr to OCPo occurring in the dark, see SI Fig.~\ref{fig:OCPr_o_solution}.

Previously published X-ray diffraction structures by Leverenz, Sutter, and co-workers \cite{Leverenz2015} show that the conjugated backbone of the bound carotenoid is twisted out of the plane of conjugation in OCPo (PDB 4XB5 \cite{4XB5}), while in OCPr N-terminal domain homologs such as red carotenoid protein (RCP) it is relatively planar (PDB 4XB4 \cite{4XB4}). The difference between the two conformations of CAN is depicted in Fig.~\ref{fig:OCP_structures_crts}b using data from X-ray diffraction structures \cite{Leverenz2015}. 
The different protein conformations containing a twisted and non-twisted form of CAN provide an uncomplicated model system to study the role of carotenoid geometry on iSF. 

\begin{figure}[!ht]
 	\centering
 	\includegraphics[scale=1.25]{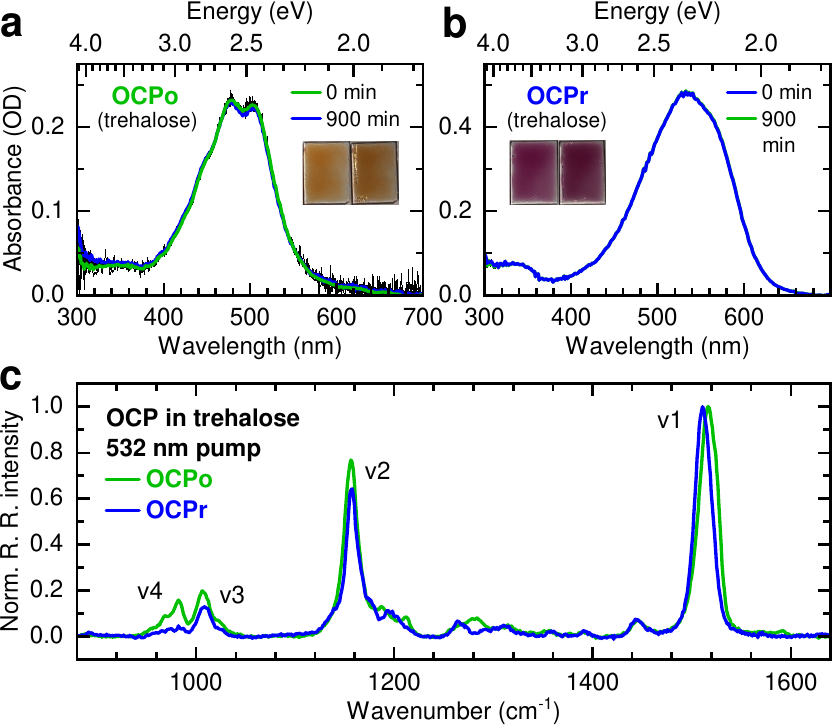}
 	\caption{\textbf{OCPo and OCPr trapped in trehalose glass films.} 
 	The steady-state absorbance spectra of the OCPo film (a) were taken in \SI{1}{\minute} intervals under constant white-light illumination (\SI{1600}{\micro\mol\photon\per\square\m\per\s}) and the spectra of the OCPr film (b) were taken in \SI{1}{\minute} intervals at \SI{22}{\degreeCelsius} in darkness. No changes in spectra were observed over \SI{900}{\min}. (c) The resonance Raman spectra of OCPo (green) and OCPr (blue) films in trehalose glass show vibrational peaks typical of carotenoids, labelled following convention. The spectra show a significant difference in the intensity of the \textnu\textsubscript{4} vibrational peak between OCPo and OCPr and a shift of the \textnu\textsubscript{1} peak.
	The Raman measurements were performed using a \SI{532}{\nano\metre} laser. Data is averaged from two successive scans, and normalized to the peak \textnu\textsubscript{1} intensity.}
 	\label{fig:OCP_steady_state}
\end{figure}

To avoid the problems associated with using spectroscopy to probe a light-activated conformational switch, we prevent the conformational change by trapping the protein in either its OCPo or OCPr conformations in a  trehalose-sucrose glass as previously described\cite{Sutherland2020}. This glass matrix prevents OCPo~$\rightleftharpoons$~OCPr conversion, as demonstrated in Fig.~\ref{fig:OCP_steady_state}(a,b), and allows us to probe each conformation in isolation at room temperature, without altering its conformation or photophysics\cite{Pidgeon2022, Kurashov2018}. 

To confirm the twisted/planar conformations of CAN in OCPo/OCPr glass films, we turn to resonance Raman spectroscopy. As described above\cite{Yu2017,Kish2015}, the presence of a so-called \textnu\textsubscript{4} peak at $\sim$\SI{980}{\per\centi\m} in the resonance Raman spectrum of carotenoids (due to out-of-plane \ce{C-H} wagging modes \cite{Saito1983}) is generally associated with a backbone twist of the carotenoid \cite{Yu2017, Kish2015}.  Fig.~\ref{fig:OCP_steady_state}c shows the resonance Raman spectrum of OCPo (blue) and OCPr (green). Consistent with previous measurements on an echinenone-binding OCP \cite{Kish2015}, we observe a larger twist-induced \textnu\textsubscript{4} peak in OCPo than in OCPr, confirming native geometry is maintained in trehalose-encapsulated OCPo and OCPr.

\begin{figure}[ht]
	\centering
	\includegraphics[scale=1.25]{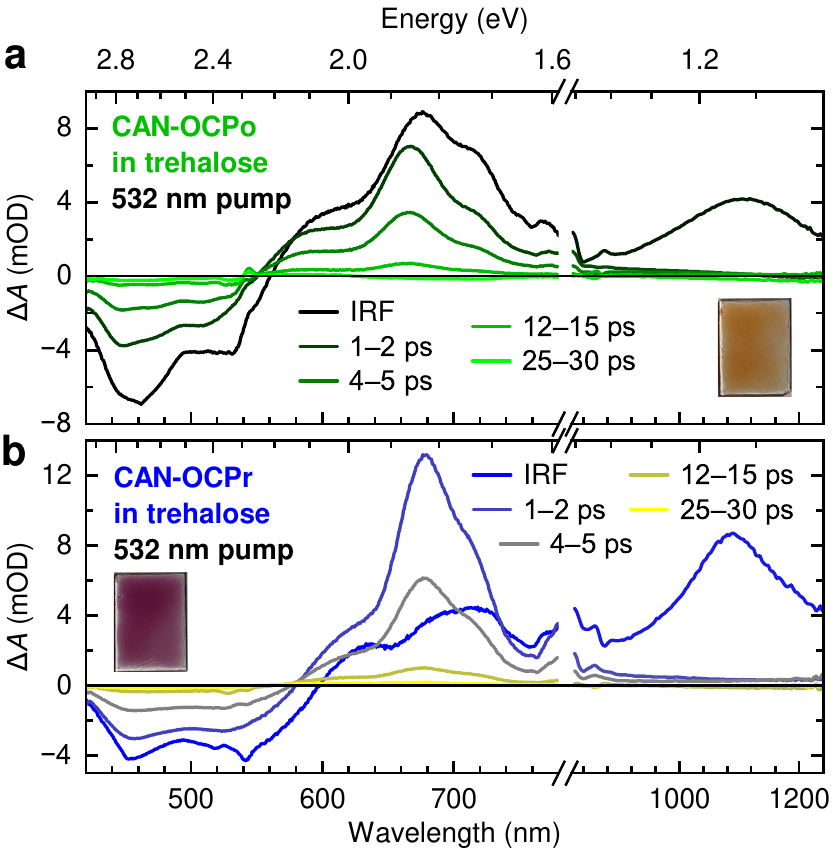}
	\caption{\textbf{Transient absorption spectra of CAN-binding OCPo (a) and OCPr (b) trapped in trehalose films.} The spectral timeslices have been averaged between the times indicated on the figure and are consistent with an S\textsubscript{2} $\rightarrow$ S\textsubscript{1} $\rightarrow$ S\textsubscript{0} decay scheme in both cases, with no discernible long-lived features (see also global lifetime analysis of the data in the SI note on Global lifetime analysis and Ref.~\cite{Pidgeon2022}). The films were excited with \SI{532}{\nano\m}, \SI{5}{\kilo\hertz}, $\sim$\SI{100}{\femto\s}, \SI{200}{\micro\J\per\square\centi\m} pump pulses.}
	\label{fig:STTA_timeslices}
\end{figure}

\begin{figure}[ht]
	\centering
	\includegraphics[scale=1.25]{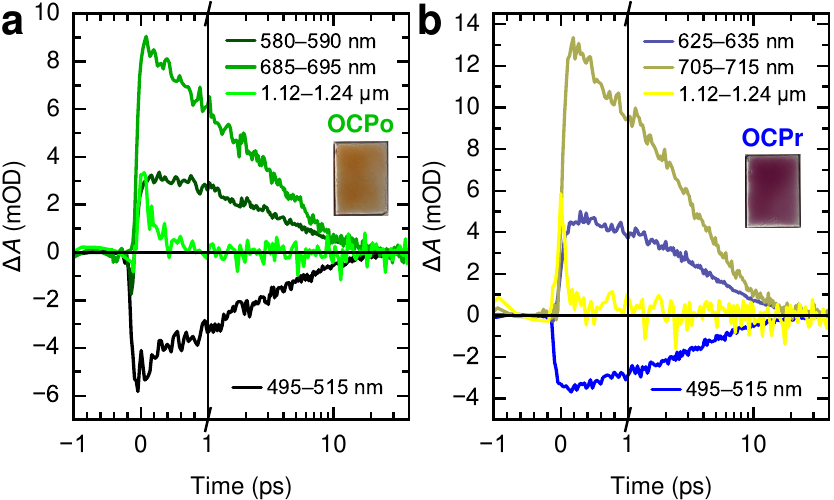}
	\caption{\textbf{Transient absorption dynamics of CAN-binding OCPo (a) and OCPr (b) in trehalose glass.} The dynamics have been averaged between the wavelengths indicated in the legend and demonstrate that no discernible long-lived features are seen. The films were excited with \SI{532}{\nano\m}, \SI{5}{\kilo\hertz}, $\sim$\SI{100}{\femto\s}, \SI{200}{\micro\J\per\square\centi\m} pulses. Note the plots have a linear time-axis up to \SI{1}{\pico\s}, and subsequently logarithmic up to \SI{40}{\pico\s}.}
	\label{fig:STTA_kinetics}
\end{figure}

Having established that the CAN backbone is more twisted in OCPo than OCPr, we test the suggestion that such a twist is the determinant for iSF reactivity\cite{Papagiannakis2002, Yu2017, Zhang2022}. Picosecond transient absorption spectroscopy spectra and dynamics are shown in Figs.~\ref{fig:STTA_timeslices} and \ref{fig:STTA_kinetics}. 
Global lifetime analysis of the data is shown in SI Fig.~\ref{fig:STTA_GLA_DADS_OCPo_532nm} and \ref{fig:STTA_GLA_DADS_OCPr_532nm}, but simply from inspection of the raw data in Fig.~\ref{fig:STTA_timeslices}, we see that all spectral features in both OCPo (green) and OCPr (blue) decay to $<$1\% of the initial population within \SI{30}{\pico\s}. Importantly, we observe no obvious formation of SF-generated triplets, as reported in light-harvesting complexes in purple bacteria \cite{Yu2017, Gradinaru2001}. Instead both OCPo and OCPr broadly demonstrate the expected isolated carotenoid behavior characterized by rapid internal conversion from S\textsubscript{2} to S\textsubscript{1} (evidenced by the instrument-limited decay of an excited-state absorption (ESA) in the near infrared region), and subsequent decay of S\textsubscript{1}-like states to the ground-state. Therefore, a twist along the carotenoid backbone is not sufficient to enable iSF. 

\section{Discussion}

The lack of intramolecular singlet fission (iSF) in the protein-twisted carotenoid in OCPo appears to counter the currently accepted explanation for singlet fission (SF) in light-harvesting complexes (LHCs) in purple bacteria  \cite{Gradinaru2001,Papagiannakis2002,Yu2017,Niedzwiedzki2017,Zhang2022}. We therefore return to the original studies of SF in these LHC systems and discuss them in light of recent work on the nature of intermediate triplet-pair states involved in singlet fission \cite{Bayliss2016, Bossanyi2021b}. 
 
\begin{figure}[ht]
	\centering
	\includegraphics[width=0.5\textwidth]{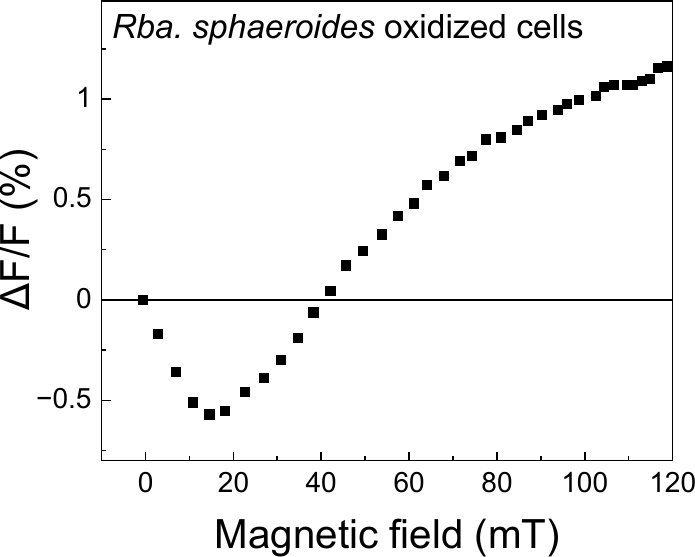}
	\caption{\textbf{Magnetic field effect (MFE) of \emph{Rhodobacter sphaeroides}.} Data reproduced from Ref.~\cite{Kingma1985a} and used with permission from Elsevier. The data is plotted as the normalized change in fluorescence $\Delta$F/F
	(detected at \SI{890}{\nano\m}) as a function of the magnetic field strength
	upon \SI{515}{\nano\m} excitation in oxidized cells of \emph{Rba.~sphaeroides} 2.4.1., with \SI{1}{\milli\molar} K\textsubscript{3}Fe(CN)\textsubscript{6} added. $A_\textrm{850--960}=0.35$; optical path length \SI{2}{\milli\m}.}
 	\label{fig:MFE}
\end{figure}

Singlet fission in LHCs was first observed in a series of experiments that probed their magnetic field-dependent fluorescence \cite{Rademaker1980, Kingma1985, Kingma1985a}. Representative data for oxidized cells from \textit{Rhodobacter sphaeroides} 2.4.1 from Ref.~\cite{Kingma1985a} are reproduced in Fig.~\ref{fig:MFE}; similar behavior has been reported for whole cells and isolated LHCs from several strains of purple bacteria  \cite{Kingma1985, Kingma1985a, Gryaznov2019, Rademaker1980}. The shape of the magnetic field effect (MFE in Fig.~\ref{fig:MFE}), with an initial dip in fluorescence as the field increases from 0 to \SI{40}{\milli\tesla}, and then a rise in fluorescence to saturation beyond \SI{100}{\milli\tesla}, is a characteristic signature of SF. 

This behavior is very well described by the kinetic model of SF by Johnson and Merrifield\cite{Merrifield1968, Groff1970, Bossanyi2021, Tapping2016, Piland2013}, published in the 1960s and 70s. 
Recent work shows \cite{Bayliss2016, Bossanyi2021b} that this low-field Merrifield-type MFE behavior can only be observed when the inter-triplet exchange interaction $J$ is negligible, or more precisely when $J\ll D$ \cite{Benk1981}, where $D$ is the intra-triplet dipolar zero-field splitting parameter. In carotenoids, and indeed most organic chromophores, $D$ is relatively small, on the order of $4$--\SI{10}{\micro\electronvolt} \cite{Frick1990, Teki1994}. If $J$ increases beyond $D$, the MFE has a different behavior, showing dips in fluorescence at much higher field strengths \cite{Bayliss2016, Bossanyi2021b, Bayliss2018, Ishikawa2018}. Therefore, to determine whether SF along a single carotenoid chain is capable of producing the measured MFEs in LHCs, we must estimate the values of $J$ and $D$. 

Before doing so we make several observations about the carotenoids involved in SF in LHCs of purple bacteria: (1) the S\textsubscript{0}$\rightarrow$S\textsubscript{2} absorbance spectra of the carotenoids in light-harvesting antenna are similar to their all-\textit{trans} forms in organic solvent and depend sensitively on carotenoid conjugation length \cite{Niedzwiedzki2017, Chynwat1995}. A full break in conjugation along the chain would lead to a dramatic blue-shift of the carotenoid absorption feature that is not observed. (2) The carotenoid T\textsubscript{1}$\rightarrow$T\textsubscript{n} excited-state absorption feature seen in transient absorption of LHCs\cite{Gradinaru2001,Papagiannakis2002, Niedzwiedzki2017, Zhang2022} is very similar to that seen in aggregated carotenoids of comparable conjugation lengths forming triplets by intermolecular SF\cite{Musser2015, Zhang2018c,Chang2017, Sutherland2020}. The T\textsubscript{1}$\rightarrow$T\textsubscript{n} feature is also sensitive to carotenoid conjugation length,\cite{Bensasson1976,Bensasson1977,Niedzwiedzki2017} and a conjugation break along the chain would similarly lead to a blue-shift that is not observed. (3) The dipolar $D$ and $E$ parameters of the SF-generated triplets in LHCs from transient electron parametric resonance (EPR) spectroscopy are similar to full-chain triplet $D$ and $E$ parameters, rather than to their half-chain alternatives \cite{Gryaznov2019}. These observations suggest that the conjugation along the chain is not broken, even in the protein, and therefore that the triplets at either end of the chain maintain orbital overlap and, presumably, non-negligible $J$.

The exchange interaction, $J$, between triplets within a pair is equal to one sixth of the energy difference between the pure singlet triplet-pair, denoted \textsuperscript{1}(TT), and the pure quintet, \textsuperscript{5}(TT). In addition, to first approximation, the energy of \textsuperscript{5}(TT) is equal to twice the free triplet energy \cite{Kollmar1993, Musser2019a, Valentine2020}. In carotenoids, as described above, the lowest energy singlet state (S\textsubscript{1}) is predominantly a pure singlet \textsuperscript{1}(TT) state. Therefore, comparison between twice the energy of a triplet on half a chain against the energy of S\textsubscript{1} on a full chain provides an indication of the exchange interaction. 

Recent high-level density matrix renormalization group (DMRG) calculations of the Pariser-Parr-Pople-Peierls
Hamiltonian\cite{Valentine2020} calculate that $2\times E(\textrm{T}_1)$ for a half chain is higher in energy than S\textsubscript{1} ($\approx$\textsuperscript{1}(TT)) for a full chain at all conjugation lengths. This is supported by experimentally determined energies: for diphenylexatriene with $N=5$ conjugated double bonds, $2\times E(\textrm{T}_1)=3.02\pm \SI{0.1}{\electronvolt}$ \cite{Bensasson1976, Chattopadhyay1983}, while for spheroidene with twice the number of double bonds ($N=10$), $E(\textrm{S}_1) = \SI{1.77}{\electronvolt}$ \cite{Fujii1998}. This would indicate an exchange interaction of $J=\SI{0.2}{\electronvolt}$, which is orders of magnitude larger than the dipolar parameter $D\sim4$--\SI{10}{\micro\electronvolt} \cite{Frick1990, Teki1994}. These energies indicate that the triplets within \textsuperscript{1}(TT) should be strongly exchange coupled.

The triplets within a single carotenoid chain are therefore exchange coupled ($J\gg D$), even in a protein that twists the carotenoid backbone \cite{Yu2017}, as no breaks in conjugation along the carotenoid chain have been observed (\textit{i.e.}~no observable shifts in absorption spectra \cite{Niedzwiedzki2017} or changes in dipolar $D$ and $E$ parameters\cite{Gryaznov2019}). Therefore, MFEs such as those reproduced in Fig.~\ref{fig:MFE}, that were the initial proof of SF in purple bacteria, cannot be explained with an intramolecular model of SF.

\section{Conclusions}

We conclude that singlet fission (SF) to produce long-lived triplets does not occur along a single twisted carotenoid chain in the OCP and is unlikely to occur in purple bacterial LHCs, contrary to the current notion \cite{Gradinaru2001, Papagiannakis2002, Niedzwiedzki2017, Yu2017, Zhang2022}. We conclude this because (1) immobilized OCPo -- an uncomplicated, minimal carotenoprotein -- shows similar twisted carotenoid geometry to LHCs, but shows no evidence of SF, and (2) the MFEs that identified SF in purple bacteria are irreconcilable with iSF without a significant break in conjugation (which is not observed). These findings therefore call into question the mechanism of SF that is observed in LHCs.

\section{Supplementary Information}

\subsection{Materials and Methods}

\subsubsection{Sample preparation}
\label{sec:methods_sample_prep}

OCP containing $\sim$100\% canthaxanthin (CAN) was produced from BL21(DE3) \emph{Escherichia coli} (\emph{E. coli}) using a dual-plasmid system comprised of pAC-CANTHipi \cite{Cunningham2007} and pET28a containing the gene encoding OCP (slr1963) from \emph{Synechocystis} sp.~PCC 6803. Briefly, \SI{500}{\milli\litre} cultures were grown at \SI{37}{\degreeCelsius} (\SI{200}{\rev\per\minute} agitation) in \SI{2}{\litre} baffled Erlenmeyer flasks using lysogeny broth (LB) medium containing the appropriate concentrations of antibiotics. When the optical density of the medium at \SI{600}{\nano\m} (OD\textsubscript{600}) had reached a value of 0.6, protein production was induced by addition of \SI{0.5}{\milli\Molar} isopropyl \textbeta-D-1-thiogalatopyranoside and the cultures incubated for \SI{16}{\hour} at \SI{18}{\degreeCelsius}. 

Cells were harvested by centrifugation (4,400$\times${}$g$, \SI{30}{\minute}, \SI{4}{\degreeCelsius}) and resuspended in binding buffer (\SI{50}{\milli\Molar} HEPES, pH 7.4, \SI{500}{\milli\Molar} NaCl, \SI{5}{\milli\Molar} imidazole). Cells were lysed by sonication and then centrifuged (53,000$\times${}$g$, \SI{30}{\minute}, \SI{4}{\degreeCelsius}). The supernatant was collected and filtered (\SI{0.22}{\micro\m} filter pores) and applied to a Chelating Sepharose Fast Flow column (GE Healthcare) pre-equilibrated with NiSO\textsubscript{4}. The column was washed with binding buffer, wash buffer (\SI{50}{\milli\Molar} HEPES, pH 7.4, \SI{500}{\milli\Molar} NaCl, \SI{50}{\milli\Molar} imidazole) and elution buffer (\SI{50}{\milli\Molar} HEPES, pH 7.4, \SI{100}{\milli\Molar} NaCl, \SI{400}{\milli\Molar} imidazole) with the elution pooled for further purification. The protein sample was buffer exchanged into buffer A (\SI{50}{\milli\Molar} HEPES, pH 7.4) loaded onto a Fast Flow Q-Sepharose column (GE Healthcare) and a linear gradient of 0--\SI{1}{\molar} NaCl was applied. Fractions were analyzed by SDS-PAGE and appropriate samples taken forward for size exclusion chromatography on a Superdex 200 Increase column (GE Healthcare) in buffer B (\SI{50}{\milli\Molar} HEPES, pH 7.4, \SI{200}{\milli\Molar} NaCl). Where necessary OCP samples were concentrated using centrifugal dialysis (VivaSpin, Sartorius).

OCPo samples were fixed in sucrose-trehalose glasses in strict darkness by mixing \SI{100}{\micro\litre} of concentrated protein solution (OD\textsubscript{max} $\sim2$) in aqueous buffer (\SI{50}{\milli\molar} HEPES, \SI{200}{\milli\molar} NaCl, pH 7.4) with \SI{100}{\micro\litre} of a trehalose-sucrose mixture (\SI{0.5}{\molar} trehalose, \SI{0.5}{\molar} sucrose). \SI{200}{\micro\litre} of the protein-trehalose mixture was drop-cast in the center of a quartz-coated glass substrate (S151, Ossila; 15$\times$20$\times$\SI{1.1}{\milli\m}). The substrate was incubated under vacuum (\SI{-70}{\kilo\pascal}) with an excess of calcium sulfate desiccant (Drierite) at room temperature for at least 48 hours. OCPr trehalose glasses were made in an identical manner, with samples illuminated for \SI{30}{\minute} (\SI{1600}{\micro\mol\photon\per\square\m\per\s}) prior to the addition of the trehalose-sucrose solution and constant weaker illumination (\SI{500}{\micro\mol\photon\per\square\m\per\s}) for the duration of the desiccation.

OCPo and OCPr samples measured with the resonance Raman setup were additionally encapsulated with imaging spacers and a cover slip to protect the trehalose against atmospheric rehydration. For these samples, a stack of two imaging spacers (SecureSeal, Grace BioLabs; \SI{9}{\milli\m} diameter, \SI{0.12}{\milli\m} thickness) were attached to the quartz-coated glass substrate (S151, Ossila; 15$\times$20$\times$\SI{1.1}{\milli\m}) and \SI{40}{\micro\litre} of the protein-trehalose mixture drop-cast in the center of the imaging spacer. The substrate was then placed in vacuum as above; pressure was released under a continuous flow of ultra-pure nitrogen gas and a glass microscope cover slip (ThermoScientific; 22$\times$\SI{22}{\milli\m}, No.1 thickness) was attached to the upper imaging spacer.

\subsubsection{Transient absorption spectroscopy}
\label{sec:methods_TA}

\begin{figure}[h]
	\centering
	\includegraphics[scale=1.25]{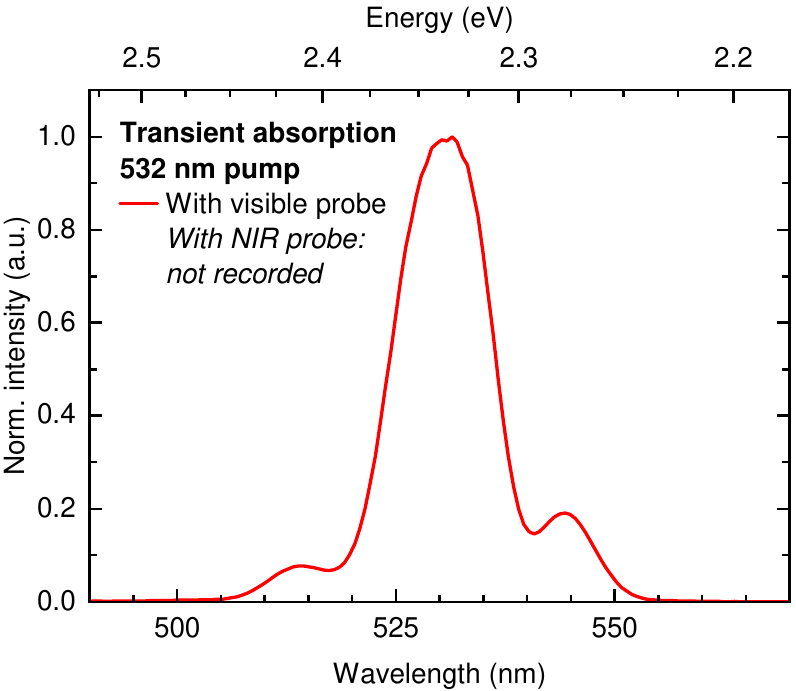}
	\caption{\textbf{Intensity spectrum of the pump used in the short-time transient absorption experiments while using the visible probe.} This \SI{532}{\nano\m}-centered (set in WinTopas4) excitation profile was used in taking the visible-probe data shown in main text. The \SI{532}{\nano\m}-centered profile used in taking the NIR-probe data was not taken, and may have had a slightly different spectrum owing to day-to-day variation in the tunable pump generation. The spectrum has been normalized to the maximum intensity.}
	\label{fig:STTA_pump_scatter}
\end{figure}

Short-time transient absorption spectroscopy was undertaken with a commercial spectrometer (Helios, Ultrafast Systems) outfitted with a Ti:Sapphire seed laser (MaiTai, Spectra Physics) providing \SI{800}{\nano\m} pulses (\SI{84}{\mega\hertz}, \SI{25}{\femto\s} nominal FWHM) and a Ti:Sapphire chirped-pulse amplifier (Spitfire Ace PA-40) amplifying \SI{800}{\nano\m} pulses (\SI{10}{\kilo\hertz}, \SI{12}{\W} average power, \SI{40}{\femto\s} nominal FWHM). Tuneable pump pulses for excitation were generated by seeding a part of the \SI{800}{\nano\m} beam in an optical parametric amplifier (TOPAS Prime, Light Conversion). An optical chopper was used to modulate the pump frequency to \SI{5}{\kilo\hertz}. An intensity spectrum for the pump used in visible-probe measurements is shown in Fig.~\ref{fig:STTA_pump_scatter}. Pump beam spot sizes were measured at the sample position with a CCD beam profiler (BC106N-VIS/M, Thorlabs), and used in subsequent calculations to tune a \SI{200}{\micro\J\per\square\centi\m} pump fluence. Supercontinuum probes were generated with a part of the \SI{800}{\nano\m} pulse focused on either a sapphire crystal for visible probes (450--\SI{800}{\nano\m}) or a YAG crystal for NIR probes (800--\SI{1600}{\nano\m}). Pump-probe delay was controlled with a motorized delay stage with a random stepping order. The signal was dispersed with a grating and detected with CMOS or InGaAs sensors for visible or NIR probes, respectively. The pump and probe polarizations were set to the magic angle. 

Surface Xplorer 4.3.0 (Ultrafast Systems) was used in processing the transient absorption datasets. Noisy edges of the spectra were trimmed, and the program’s bad spectra replacement procedure was applied. A background correction (`subtract scattered light') was then applied using the spectra before any apparent response from the sample. Chirp correction was applied, choosing points at the first apparent signal for a given kinetic. Time zero is adjusted to the time of maximum initial signal. Further processing and some analysis was performed with original Python code.

\subsubsection{Resonance Raman spectroscopy}

Resonance Raman measurements were performed with a Renishaw inVia Raman system (Renishaw plc., Wotton-Under-Edge, UK) in a backscattering configuration. A \SI{532}{\nano\m} laser (150--\SI{750}{\micro\watt} power) and a 50$\times$ objective were used (NA 0.50, spot size $\sim$\SI{1}{\micro\metre}). Acquisition times used were in the range 5--\SI{30}{\second}.

\subsection{Figure preparation}

OriginPro 9.6.0.172 (OriginLab) and home-built Python code was used to prepare the plots. 

\clearpage

\subsection{Additional steady-state absorbance data}

\clearpage

\subsection{Additional resonance Raman data}

\begin{figure}[h]
	\centering
	\includegraphics[scale=1.25]{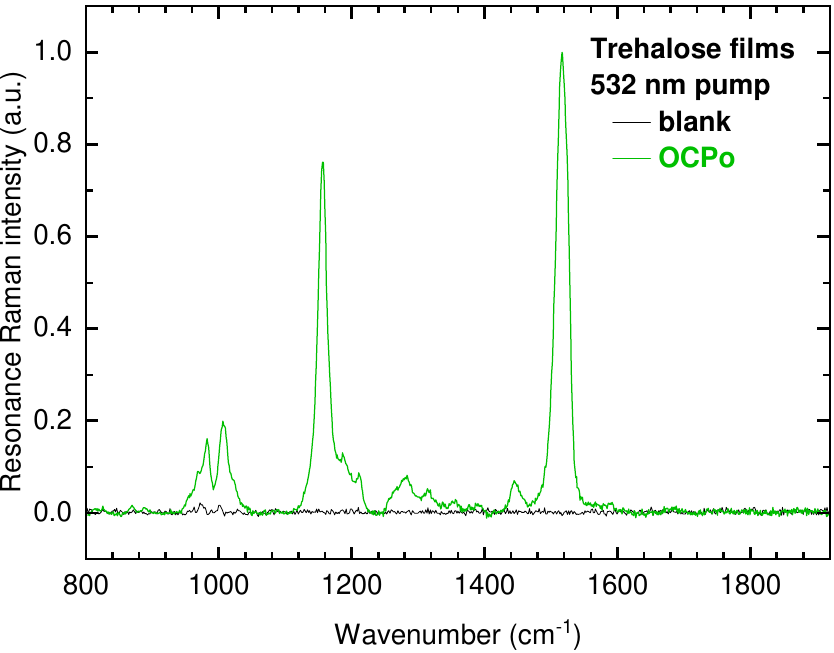}
	\caption{\textbf{Resonance Raman spectra of blank trehalose glass and OCPo in trehalose.} The blank trehalose spectrum (black) is almost entirely noise, confirming that the trehalose-sucrose glass is not contributing significant Raman signal in any of the samples incorporating trehalose glass. OCPo in trehalose (green) is shown as a comparison, normalized to the peak \textnu\textsubscript{1} intensity. The blank spectrum has been scaled to give a similar noise magnitude at the high-wavenumber end. Both spectra are single scans (not averaged). Measurements were undertaken at room temperature. A \SI{532}{\nano\metre} pump was used.}
	\label{fig:trehalose_resonance_Raman}
\end{figure}

\clearpage

\subsection{Additional transient absorption data}

\subsubsection{Pump spectrum}

\clearpage

\subsubsection{Additional note: global lifetime analysis}
\label{sec:GLA}

Global lifetime analysis on the visible-probe transient absorption data of OCPo and OCPr was performed. This was done using the Glotaran 1.5.1 software package\linebreak (http://glotaran.org) \cite{Snellenburg2012}, a GUI for the R package TIMP\cite{Mullen2007}. Data used had already been processed with the steps outlined in Section \ref{sec:methods_TA}; in particular, a chirp correction had already been applied, so that a term to account for chirp did not need to be included in the fitting. Noisy regions in the data due to pump scatter were excluded for all times to ensure a good fit of the rest of the data. Noisy red and blue ends in the data associated with tails of the probe were also excluded, so that the fitted wavelengths were \SI{430}{\nano\m} to \SI{780}{\nano\m}. The fitting was weighted favourably at later delay times for good fits of any long-lived features; Table \ref{tab:STTA_GLA_DADS_weighting} shows the weighting applied. Due to the strong coherent artifact feature in the first \SI{0.5}{\pico\s}, only data beyond that time was fitted. Thus, in the model, terms to account for the coherent artifact and S\textsubscript{2} states were not included. This left a relatively simple fitted model of a number of wavelength-dependent decay-associated difference spectra (DADS) decaying exponentially in parallel.

\begin{table}[ht]
	\renewcommand*{\arraystretch}{1.2} 
	\newcolumntype{C}[1]{>{\centering\arraybackslash}m{#1}} 
	\begin{center}
		\small 
		\begin{tabular}{|C{38mm}|C{27mm}|}
			\hline
			\textbf{Time range (\SI{}{\pico\s})} & \textbf{Weighting} \\
			\hline
			0.5~--~10 & 1 \\ \hline
			10~--~20 & 2 \\ \hline
			20~--~30 & 3 \\ \hline
			30~--~35 & 4 \\ \hline
			$>$35 & 5 \\ \hline
		\end{tabular}
	\end{center}
	\vspace{-6pt}
	\caption{\textbf{Weightings applied to time-ranges of the visible \SI{}{\pico\s} transient absorption data for the global lifetime analysis.} Note that only data $>$\SI{0.5}{\pico\s} was fitted, and that the maximum time delay in these experiments was $\sim$\SI{40}{\pico\s}.}
	\label{tab:STTA_GLA_DADS_weighting}
\end{table}

\begin{figure}[ht]
	\centering
	\includegraphics[scale=0.8]{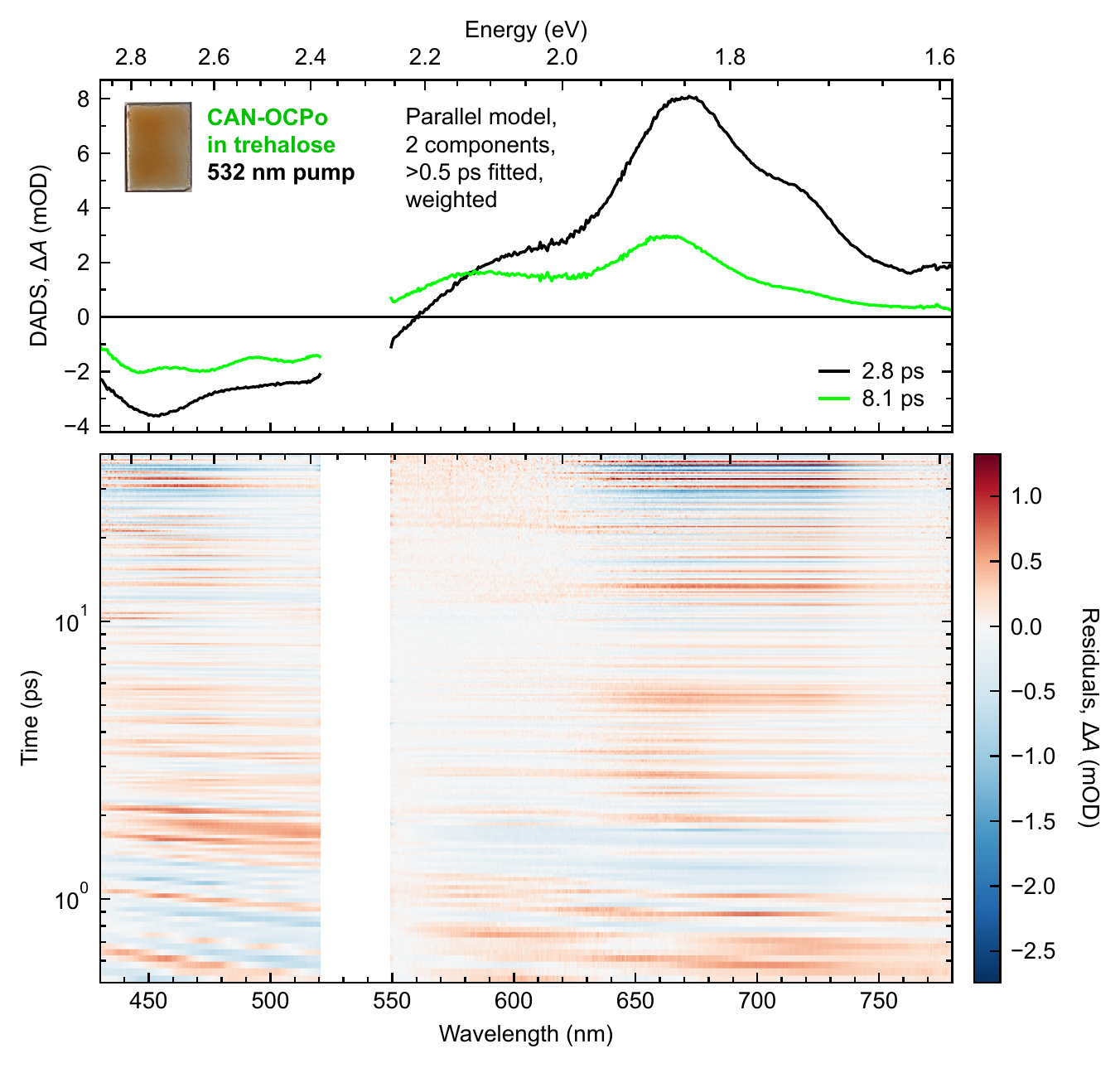}
	\caption{\textbf{Results of global lifetime analysis with a 2-component parallel model on transient absorption data of CAN-binding OCPo in trehalose with peak pump wavelength \SI{532}{\nano\m} and a visible probe: DADS (top) and residuals (bottom).} Only the wavelength range 430--\SI{780}{\nano\m} and times $>$\SI{0.5}{\pico\s} were fitted, and noisy data from 520.5--\SI{549.5}{\nano\m} due to significant pump scatter was excluded from the fit. DADS lifetimes are specified in the legend. $\textrm{Residuals} = \textrm{Data}-\textrm{Fit}$. See text for further details.}
	\label{fig:STTA_GLA_DADS_OCPo_532nm}
\end{figure}

\begin{figure}[ht]
	\centering
	\includegraphics[scale=0.8]{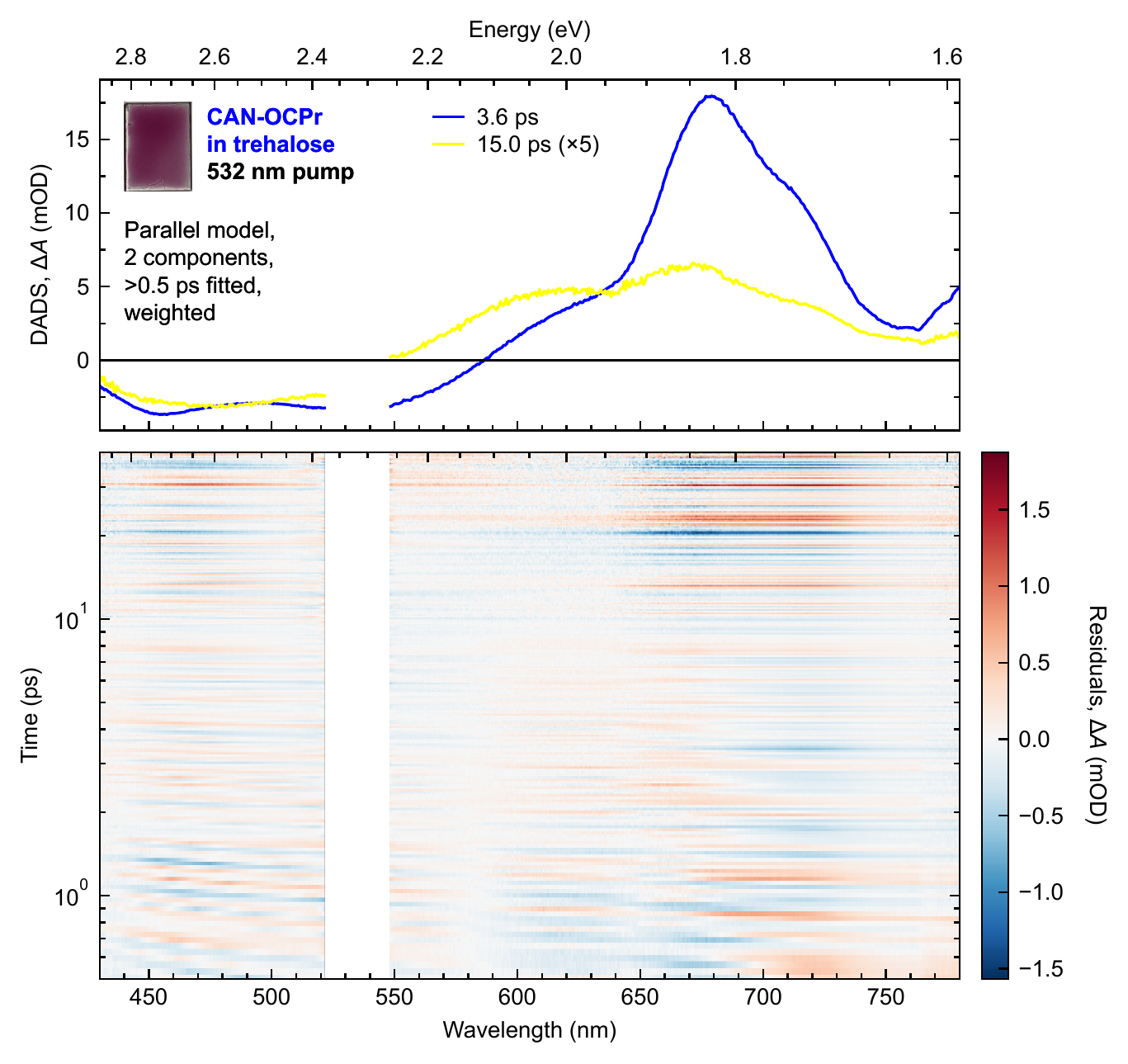}
	\caption{\textbf{Results of global lifetime analysis with a 2-component parallel model on transient absorption data of CAN-binding OCPr in trehalose with peak pump wavelength \SI{532}{\nano\m} and a visible probe: DADS (top) and residuals (bottom).} Only the wavelength range 430--\SI{780}{\nano\m} and times $>$\SI{0.5}{\pico\s} were fitted, and noisy data from 522--\SI{548}{\nano\m} due to significant pump scatter was excluded from the fit. DADS lifetimes are specified in the legend; multiplications refer to scalings applied to the DADS. $\textrm{Residuals} = \textrm{Data}-\textrm{Fit}$. See text for further details.}
	\label{fig:STTA_GLA_DADS_OCPr_532nm}
\end{figure}

Global lifetime analysis of the OCP data does not indicate any significant population of longer-lived features. 2-component global lifetime analysis of the CAN-binding OCPo and OCPr data are shown in Fig.~\ref{fig:STTA_GLA_DADS_OCPo_532nm} and \ref{fig:STTA_GLA_DADS_OCPr_532nm}, respectively. Fitting a 2-component parallel decay model in an artifact-free region of the visible-probe data beyond the initial coherent artifact and S\textsubscript{2}-associated response gives two decay-associated difference spectra (DADS) for both the OCPo data and OCPr data, with the longer time-constant DADS relatively weaker and blueshifted in both cases. Both components are likely associated with mixed S\textsubscript{1} decay, internal vibrational redistribution, and vibrational cooling.\cite{Balevicius2019} A single component is not sufficient to adequately fit the region of the data, and a fitting a third component gives results with spurious DADS profiles.

We note that sample degradation (caused the $\sim$\SI{200}{\micro\J\per\square\centi\m} pump fluence) likely affects the fitted time constants and DADS profiles, and that the maximum time delay used was about \SI{40}{\pico\s}.

\begin{acknowledgement}


The authors thank James D.~Shipp, Sayantan Bhattacharya and David G.~Bossanyi for assistance with transient absorption measurements.


\begin{sloppypar}
\textbf{Funding:} G.A.S.~and C.N.H.~acknowledge ERC Synergy Grant 854126. J.P.P.~thanks the EPSRC for support through a Doctoral Training Partnership Scholarship (EP/R513313/1). The authors thank the EPSRC for a Capital Equipment Award (EP/L022613/1 and EP/R042802/1) which funded the Lord Porter Laser facility used in this study. J.C., C.N.H, G.A.S.~and S.W.~thank the EPSRC for funding through EP/S002103/1. J.C.~and S.W.~also thank the EPSRC for funding through EP/N014022/1. M.S.P.~and M.P.J.~were supported by Leverhulme Trust award RPG-2019-045. A.H.~acknowledges The Royal Society (award URF{\textbackslash}R1{\textbackslash}191548). H.K.H.L.~and W.C.T.~acknowledge the SPECIFIC Innovation and Knowledge Centre (EP/N020863/1) grant for providing financial support.

\textbf{Author contributions:} G.A.S.~conceived the study. G.A.S., J.P.P., S.W.~and J.C.~designed the experiments. G.A.S., M.S.P.~and A.H.~prepared protein samples under the supervision of M.P.J.~and C.N.H.~~Steady-state measurements were performed by G.A.S. Resonance Raman measurements were performed by H.K.H.L.~under the supervision of W.C.T.~~Transient absorption was conducted by J.P.P.~and S.W.~ within the Lord Porter Laser, with D.C. providing facility management. ~~J.P.P.~and G.A.S.~analyzed the data. J.P.P., G.A.S.~and J.C.~wrote the manuscript and prepared the figures with input from all authors.
\end{sloppypar}

\end{acknowledgement}

\begin{suppinfo}


\begin{itemize}
\item \textbf{Supplementary information (SI):} Canthaxanthin-binding OCP production, OCPo and OCPr in sugar glasses preparation, spectroscopic setups, and data analysis procedures are detailed. A figure showing back-conversion of solution OCPr to OCPo in the dark is included. Figures showing a short-time transient absorption pump spectrum and a resonance Raman spectrum of blank trehalose are included. The results of global lifetime analysis on the transient absorption data is detailed.
\item Data will be available on Sheffield University's repository, ORDA, once the paper has been accepted for publication.
\end{itemize}

\end{suppinfo}

\newpage
\clearpage
\bibliography{library}

\end{document}